\documentclass[prd,twocolumn,aps,psfig,showpacs,nofootinbib,nobibnotes,superscriptaddress,preprintnumbers,times]{revtex4}
\usepackage{graphicx,bm,color}
\graphicspath{{./fig/}{./png/}}

\newcommand{\EQ}{\begin{equation}}
\newcommand{\EN}{\end{equation}}
\newcommand{\EQA}{\begin{eqnarray}}
\newcommand{\ENA}{\end{eqnarray}}

\newcommand{\Fig}[1]{Fig.~\ref{#1}}

\newcommand{\Tab}[1]{Table~\ref{#1}}

{}
{}
{}

\def\epsK{\epsilon_{\rm K}}
\def\epsM{\epsilon_{\rm M}}

\def\EEK{{\cal E}_{\rm K}}
\def\EEM{{\cal E}_{\rm M}}

\def\EEGW{{\cal E}_{\rm GW}}

\def\OmGW{{\Omega}_{\rm GW}}
\def\hrms{{h}_{\rm rms}}

\newcommand{\GeV}{\,{\rm GeV}}

\newcommand{\ygafd}[4]{, #4, Geophys. Astrophys. Fluid Dyn. {\bf #2}, #3 (#1).}

\begin{document}

\title{Circular polarization of gravitational waves from early-Universe helical turbulence}

\date{\today}
\preprint{NORDITA-2020-102}

\author{Tina~Kahniashvili}
\email{tinatin@andrew.cmu.edu}
\affiliation{McWilliams Center for Cosmology and Department of Physics, Carnegie Mellon University, Pittsburgh, PA 15213, USA} 
\affiliation{Faculty of Natural Sciences and Medicine, Ilia State University, 0194 Tbilisi, Georgia} 
\affiliation{Abastumani Astrophysical Observatory, Tbilisi, GE-0179, Georgia}
\affiliation{Department of Physics, Laurentian University, Sudbury, ON P3E 2C, Canada}

\author{Axel~Brandenburg}
\email{brandenb@nordita.org}
\affiliation{Nordita, KTH Royal Institute of Technology and Stockholm University, 10691 Stockholm, Sweden}
\affiliation{Department of Astronomy, AlbaNova University Center, Stockholm University, 10691 Stockholm, Sweden} 
\affiliation{Faculty of Natural Sciences and Medicine, Ilia State University, 0194 Tbilisi, Georgia} 
\affiliation{McWilliams Center for Cosmology and Department of Physics, Carnegie Mellon University, Pittsburgh, PA 15213, USA}

\author{Grigol~Gogoberidze}
\email{Grigol_Gogoberidze@iliauni.edu.ge}
\affiliation{Faculty of Natural Sciences and Medicine, Ilia State University, 0194 Tbilisi, Georgia}

\author{Sayan~Mandal}
\email{sayan.mandal@stonybrook.edu}
\affiliation{Physics and Astronomy Department, Stony Brook University, Stony Brook, New York 11794, USA} 
\affiliation{Faculty of Natural Sciences and Medicine, Ilia State University, 0194 Tbilisi, Georgia}

\author{Alberto~Roper~Pol}
\email{roperpol@apc.in2p3.fr}
\affiliation{Universit\'e de Paris, CNRS, Astrophysique et Cosmologie, Paris, F-75013, France} 
\affiliation{Faculty of Natural Sciences and Medicine, Ilia State University, 0194 Tbilisi, Georgia}

\begin{abstract}
We perform direct numerical simulations to compute the net circular
polarization of gravitational waves from helical (chiral) turbulent 
sources in the early Universe for a variety of initial conditions, 
including driven (stationary) and decaying turbulence.
We investigate the resulting gravitational wave signal assuming different 
turbulent geneses such as magnetically or kinetically driven cases.
Under realistic physical conditions in the early Universe we compute 
numerically the wave number-dependent
polarization degree of the gravitational waves. 
We find that the spectral polarization degree
strongly depends on the initial conditions. 
The peak of the spectral polarization degree occurs 
at twice the typical wavenumber of the source, as expected,
and for fully helical decaying turbulence, it reaches
its maximum of nearly 100\% {\it only} at the peak.
We determine the temporal evolution of the turbulent sources as well as
the resulting gravitational waves, showing that the dominant contribution 
to their spectral energy density happens shortly after the activation of the source.
Only through an artificially prolonged decay of the turbulence can further
increase of the gravitational wave amplitude be achieved. 
We estimate the detection prospects for the net polarization, arguing that
its detection contains {\it clean} information (including the generation 
mechanisms, time, and strength) about the sources of possible parity
violations in the early Universe.
\end{abstract}

\maketitle

\section{Introduction}
A remarkable possible source of stochastic gravitational waves (GWs)
from the early Universe is turbulence in the primordial plasma 
induced either from cosmological first-order (electroweak or QCD) phase transitions
\cite{Witten:1984rs,Hogan:1986qda,Kamionkowski:1993fg},
or from the primordial magnetic fields that are coupled to the cosmological plasma \cite{Brandenburg:1996fc,Christensson:2000sp,Kahniashvili:2010gp,Brandenburg:2017rnt}.
The GW signal is potentially detectable 
by the Laser Interferometer Space Antenna (LISA) in the case of
strong enough turbulent motions present at the electroweak scale
(assuming that the total energy in the turbulence is up to 1--10\% 
of the total thermal energy at the moment of generation)
\cite{Kosowsky:2001xp,Caprini:2006jb,Gogoberidze:2007an,Kahniashvili:2008pe,Caprini:2009yp}. 
Since GWs propagate {\it almost} freely from the moment of generation 
until today\footnote{We discard the GW damping due to neutrino free 
streaming \cite{Durrer:1997ta,Weinberg:2003ur} or from anisotropic
stresses \cite{Deryagin:1986qq}.}, the detection of GWs sourced by 
primordial turbulence will open a new window to understand physical 
processes in the very early stages of the evolution of the Universe (at the 
femtoseconds timescale); see Ref.~\cite{Caprini:2019egz}
and references therein.
Moreover, several theoretical extensions of the standard model (SM)
of particle physics and cosmology (which is insufficient to explain the
matter-antimatter asymmetry in the Universe) imply
parity symmetry\footnote{
The baryon asymmetry of the Universe can be explained 
through spontaneous lepton number symmetry breaking at a cosmological phase
transition \cite{Cohen:1990it,Cohen:1990py}.} 
violation at the electroweak energy scale being
possibly manifested through helical (chiral) turbulent motions and/or
magnetic fields \cite{Long:2013tha,Dorsch:2016nrg}.
As expected, such parity-violating turbulent
sources will produce circularly polarized GWs 
\cite{Kahniashvili:2005qi,Kahniashvili:2008er,Kisslinger:2015hua,
Alexander:2018fjp,Anand:2018mgf,Niksa:2018ofa,Ellis:2020uid}
analogously to the GWs produced via Chern-Simons coupling
\cite{Alexander:2004us,Lyth:2005jf}.
Furthermore, chiral inflationary GWs might be responsible (through the  
gravitational anomaly) for gravitational leptogenesis which, in turn, manifests itself 
in the neutrino sector by being successful in either the Dirac or Majorana
neutrino mass scenario \cite{Adshead:2017znw}.
Moreover, the gravitational anomaly effect and, correspondingly, a 
successful electroweak baryogenesis scenario appear viable in
the case when the GWs are generated through helical magnetic fields
\cite{Abedi:2018top}.
Additionally, the primordial magnetic fields that are responsible for
the circularly polarized GWs and, correspondingly, for the baryon asymmetry,
can serve as seeds for large-scale magnetic fields in the Universe;
see Ref.~\cite{Vachaspati:2020blt} for a review and references therein.
Under this consideration the scheme is as follows: helical magnetic fields 
act as seeds for large-scale magnetic fields, produce circularly polarized
GWs, and enable successful lepto- and baryogenesis through the gravitational anomaly.
Obviously, if detected, the GW polarization can be a {\it
clean} measure of the deviations from the SM and will provide us with a
better understanding of the nature of parity symmetry and its violation.
In fact, the consideration of the different forms of ``driving'' 
is dictated by different possible sources: for scenarios in which
the magnetic field is the primary (dominant) source of turbulence 
(e.g., inflationary or phase transition-generated magnetic fields) we deal
with ``magnetically driven'' turbulence, while turbulence with primary 
fluid motions (e.g., sound waves) is ``kinetically driven'', and could 
then result in magnetic field generation through dynamo action. 
This can lead to a growth of the magnetic energy by orders of
magnitude even though the kinetic energy of the turbulence decays
\cite{Brandenburg:2017rnt}.
However, this is not considered in the present paper.

Each of these sources might be parity symmetry violating ones.
As we show below, the polarization spectra are different for different
forms of driving, and thus the measurement of these spectra can lead to
an identification of the source and the energy-scale at which parity
violation happened. 
Interestingly, our results may help us understand the origin of the
cosmic seed magnetic field, i.e., it may help discriminate between
astrophysical and cosmological magnetogenesis scenarios.
Perhaps most importantly, the investigation of the GW background in 
combination with its polarization might convincingly establish
that observations demand a cosmological magnetic field that cannot be 
generated by a mechanism operating within the confines of the SM.

The detection of circular polarization of the 
stochastic GW 
background is a challenging task \cite{Seto:2006hf,Smith:2016jqs,Masui:2017fzw}, and
the planar interferometers cannot measure net polarization in the case
of isotropic backgrounds \cite{Seto:2007tn,Seto:2008sr}.
However, the dipolar anisotropy induced by our proper motion
with respect to the cosmic reference frame makes it possible to
measure the net circular polarization of the stochastic GW background
\cite{Seto:2006dz,Crowder:2012ik}, and recently it has been shown that 
the net polarization of GWs could be detected with a signal-to-noise ratio 
of order one by LISA if the strength of the signal achieves $h_0^2
\Omega_{\rm GW}\sim 10^{-11}$ (with $\Omega_{\rm GW}$ the fraction
between the GW energy density and the critical
density today ${\mathcal E}_{\rm cr}=3H_0^2/(8\pi G)$, with
$H_0=100 h_0\,{\rm km}\,{\rm s}^{-1} {\rm Mpc}^{-1}$ the Hubble parameter today and
$G$ is the gravitational constant) \cite{Domcke:2019zls}. 
These findings make it extremely important to properly compute all 
characteristics (such as the amplitude, the spectral shape, and the
polarization degree) of the GW signal from primordial helical (chiral)
sources.

When computing the GW signal from early-Universe turbulent sources,
previous studies (with the exception of Ref.~\cite{Pol:2019yex}) assumed
stationary hydrodynamic turbulence and a turbulence duration set by a
fraction of the Hubble time at the moment of generation ($H_\star^{-1}$),
making it possible to use the simplified GW equation with a discarded
term that describes the expansion of the Universe 
($\sim H=a^{-1}{da/dt_{\rm ph}}$, where $t_{\rm ph}$ denotes the physical
time, and $a$ is the scale factor). 
Moreover, the magnetic field and primordial plasma coupling, and 
correspondingly, the turbulence decay have been neglected in those analytic studies, making it
impossible to study the GW source dynamics and the temporal dependence
of the GW amplitude and spectral characteristics beyond the dilution due
to the expansion of the Universe.
These shortcomings will be avoided by numerically simulating
nonstationary turbulence with all terms included, allowing for a
full coupling between magnetic fields and plasma motions. 
We apply hydrodynamic and magnetic forcing terms that have a realistic
representation of the time dependent generation of turbulence.

\begin{figure*}[t!]\begin{center}
\includegraphics[width=\textwidth]{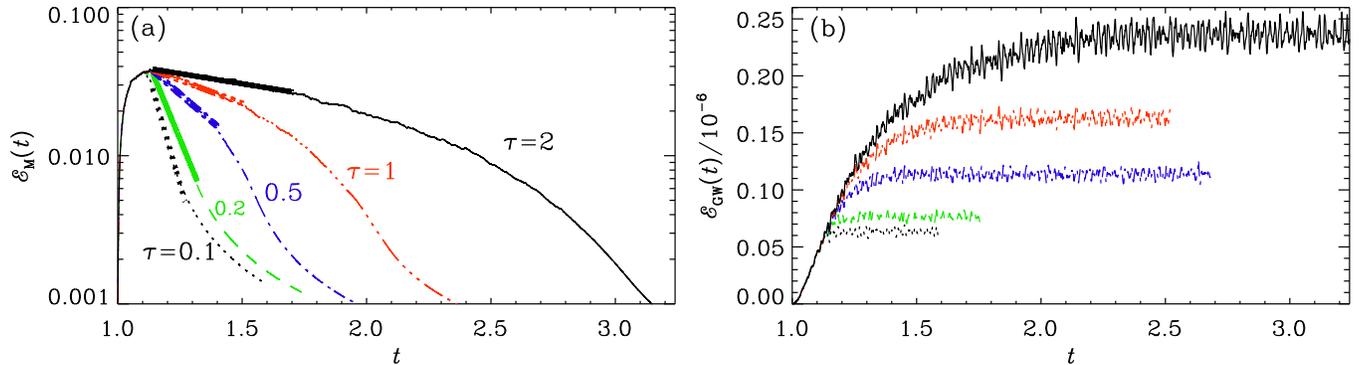}
\end{center}\caption[]{\small
Evolution of magnetic energy (left) and growth of GW energy density
(right) for simulations where the driving is turned off at $t=1.1$ 
(black dotted line), or the strength of the driving is reduced linearly
in time over the duration $\tau=0.2$ (green),  $0.5$ (blue),
$1$ (red), or $2$ (black). 
Time is in units of the Hubble time at the moment of source activation.
The magnetic and GW energy densities are
in units of the radiation energy density.
}\label{pe_comp}\end{figure*}

\section{Numerical Modeling}

Recently we have performed for the first time direct numerical simulations
of magnetohydrodynamic (MHD) turbulence in the early Universe accounting
for the expansion of the Universe using the {\sc Pencil Code} \cite{PC},
and numerically computed the resulting stochastic GW background and
relic magnetic fields \cite{Pol:2019yex}. 
To scale out expansion effects, we use appropriately scaled comoving 
variables and conformal time.
The full set of MHD equations is then similar to the usual MHD equations
\citep{Brandenburg:1996fc}, and the GW equation is written for the scaled
strains and the comoving total (magnetic and kinetic) 
traceless-transverse 
stress tensor $T_{ij}^{\rm TT}$.
These simulations allowed us to conclude that the proper inclusion of
coupling effects results in the extension of the GW spectrum at 
lower frequencies 
due to the power transfer at large scales,
until the causal horizon determined by the comoving Hubble frequency,
$f_H=1.65 \cdot 10^{-5}{\rm Hz} \, (g_\star/100)^{1/6} (T_\star/100{\rm GeV})$
with $g_\star$ and $T_\star$ being the relativistic degrees of freedom
and the temperature at the moment of the GW source activation.
In terms of the GW energy density per logarithmic frequency interval
as a fraction of critical density today, $h_0^2\Omega_{\rm GW}(f)$, the
spectrum in the frequency range $f_{\rm H} \leq f \leq f_{\rm S}$ (where 
the typical frequency of the source is 
$f_{\rm S}=2 N f_{\rm H}$\footnote{The factor ``2'' is due to the quadratic nature of
the turbulent source.}, and $N$ determines the number of turbulent 
eddies per linear Hubble length scale)\footnote{The parameter 
$N$ is determined by the physical stirring wave number of the source, 
$k_0^{\rm phys}$, via $H_\star/N =2\pi/k_0^{\rm phys}$, and to the
comoving peak wave number through $k_0=k_0^{\rm phys} (a_\star/a_0)$.
Here $a_\star$ and $a_0$ are scale factors corresponding to the 
moment of generation and today, and $a_\star/a_0 = 8.0 \times 10^{-16}
(100/g_\star)^{1/3}(100\,{\rm GeV}/T_\star)$.} is $\propto f$, as
opposed to $f^{3}$ for the causal low-frequency ($f<f_H$) tail obtained analytically
\cite{Kahniashvili:2008pe}, 
while in terms of the {\it comoving} dimensionless strain
amplitude $h_c$, conventionally written as
$h_c(f) = 1.263 \cdot 10^{-18} (f/1{\rm \, Hz})^{-1} \left [h_0 \Omega_{\rm 
GW}(f)\right]^{1/2}$ \cite{Romano:2016dpx},
we observe the scaling $h_c \propto f^{-1/2}$
in the frequency range $f_H < f < f_S$.
At this point we distinguish three parts of
$\Omega_{\rm GW}(f)$: the low-frequency region
below the causal horizon $f<f_H$, 
the intermediate region $f_H < f < f_S$, and 
the high-frequency region $f>f_S$. 
However, due to computational limitations, we are unable
to reproduce the entire spectrum in our numerical simulations. 
At high frequencies, $f>f_{\rm S}$, numerical simulations agree 
well with the analytical estimate.
In particular, for Kolmogorov turbulence with a slope $-5/3$,
the high frequency tail $h_c(f)$ ($\Omega_{\rm GW}(f)$) scales as 
$\propto f^{-7/3}$ ($\propto f^{-8/3}$); see Ref.~\cite{Pol:2019yex} for 
more details.
Earlier work \cite{Niksa:2018ofa} showed that the high-frequency
scaling of the GW spectrum depends strongly on the assumptions about
the turbulence and the modeling of the time-decorrelation function.
Since the spectrum of the stochastic GW 
background determines the detectability of GWs in a given experiment, realistic
turbulence simulations are essential for establishing the sensitivity of 
upcoming GW experiments to early-Universe physics \cite{Romano:2016dpx}.

\begin{figure*}[t!]\begin{center} 
\includegraphics[width=\textwidth]{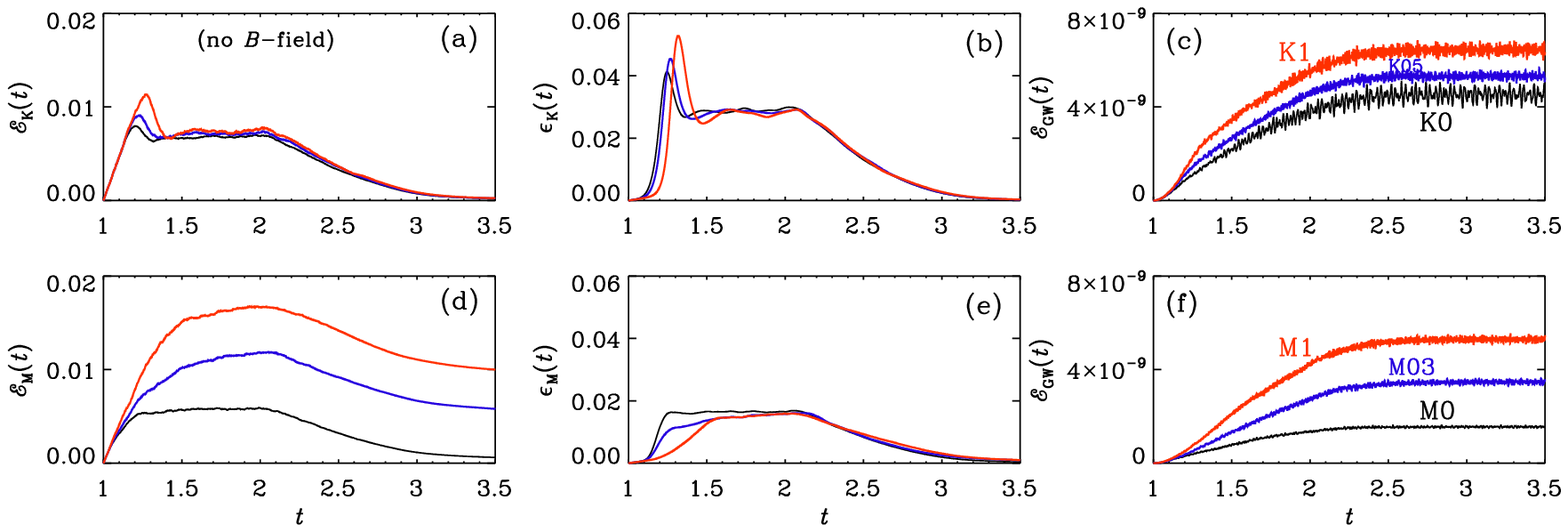}
\end{center}\caption[]{\small 
Evolution of (a) $\EEK$, (b) $\epsK$, and (c) $\EEGW$ for kinetically
driven cases with $\sigma=0$ (Run~K0, black), $0.5$ (Run~K05, blue),
and $1$ (Run~K1, red),
and of (d) $\EEM$, (e) $\epsM$, and (f) $\EEGW$ for magnetically
driven cases with $\sigma=0$ (Run~M0, black), $0.3$ (Run~M03, blue),
and $1$ (Run~M1, red).
}\label{pcomp_M512k2}\end{figure*}

\section{Results}

The spectral form of the GW spectrum at low frequencies
is independent of the initial conditions and the turbulence model while
the amplitude of the signal strongly depends on the model chosen. 
The universal form of the GW spectrum does not allow us to discriminate
between helical and nonhelical sources and thus limits our ability to
determine the parity violation in the early Universe. 
This leads us to the present study with its main focus on GW circular
polarization estimates and the question whether the detection of
polarization can help in the identification of distinct properties of
the source.

\begin{table}[b]\caption{
Characteristic parameters of the runs of Fig.~\ref{pe_comp}.
}\vspace{12pt}\centerline{\begin{tabular}{cccccc}
Run & ${\cal P}$ & $\EEM^{\max}$ & $\EEK^{\max}$ & $\EEGW^{\rm sat}$ & $\hrms^{\rm sat}$\\
\hline
$\lambda=0.1$ & $-13.5$ & $0.0367$ & $0.0134$ & $6.3\times10^{-8}$ & $2.5\times10^{-6}$\\
$\lambda=0.2$ & $-9.37$ & $0.0368$ & $0.0135$ & $7.6\times10^{-8}$ & $2.8\times10^{-6}$\\
$\lambda=0.5$ & $-3.32$ & $0.0372$ & $0.0137$ & $1.1\times10^{-7}$ & $3.4\times10^{-6}$\\
$\lambda=1$   & $-1.43$ & $0.0378$ & $0.0140$ & $1.6\times10^{-7}$ & $4.1\times10^{-6}$\\
$\lambda=2$   & $-0.63$ & $0.0381$ & $0.0141$ & $2.4\times10^{-7}$ & $5.1\times10^{-6}$\\
\label{Tsummary_ramp}\end{tabular}}\end{table}

\subsection{Decay with Decreased Driving}

We first address the temporal evolution of the GW spectrum.
As it was shown in Ref.~\cite{Pol:2019yex}, the GW spectrum becomes
stationary shortly after the driving of the source ends (i.e.,
when the free decay stage of the source starts),
while the energy density of the source is still present. 
To demonstrate this, we drive magnetic fields with an electromotive force, 
${\cal F}_i=(\delta_{ij}+\sigma\epsilon_{ijl}\hat{k}_l){\cal F}_j^{(0)}$,
consisting of plane waves that are delta-correlated in time.
Here, $-1\leq\sigma\leq1$ quantifies the fractional helicity,
and ${\cal F}_j^{(0)}$ is a nonhelical plane-wave forcing term.
Plasma motions are selfconsistently driven by the Lorentz force.
Purely hydrodynamic motions are driven by a ponderomotive force
analogous to ${\cal F}_i$.

In Fig.~\ref{pe_comp}, we show the temporal 
evolution of the source and the growth of the GW energy density 
for the driven ($1<t<1.1$) and decaying stages ($t>1.1$), where 
the driving decreases linearly for a duration $\tau=0.1$--$2$,
although $\tau>0.5$ may be unrealistic. 
As in Ref.~\cite{Pol:2019yex}, we use $1152^3$ meshpoints
for the runs in Fig.~\ref{pe_comp} and put $\sigma=0$;
see \Tab{Tsummary_ramp} for a summary relevant quantities.
During the statistically stationary stage, the GW energy density growth rate is 
proportional to the duration of turbulence, as was estimated 
through analytical modeling of 
Ref.~\cite{Gogoberidze:2007an}.
In reality, the driving stage is short compared to the 
Hubble time-scale, and consists of the few largest eddy turnover times.

In \Tab{Tsummary_ramp}, we have quoted the values of $\EEGW^{\rm sat}$ and
$\hrms^{\rm sat}$ obtained at the end of the simulation at $t=t_{\rm end}$.
We emphasize that $\EEGW$ is the comoving GW energy density
normalized by the critical energy density, 
which is the same as the radiation energy density during
the simulation, and $\hrms$ corresponds to the scaled strain.
To compute the relic observable $h_0^2\OmGW$ at the present time,
we have to multiply $\EEGW^{\rm sat}$ by a factor
$(H_\ast/H_0)^2 a_0^{-4}$;
see Refs.~\cite{Pol:2019yex,Pol:2018pao} for details,
although there we used the symbol $\OmGW$ also for the latter.
For the aforementioned fiducial parameters of $g_*=100$ and
$T_*=100\GeV$, this factor is $1.64\times10^{-5}$.
The largest value of $\EEGW^{\rm sat}$ quoted in
\Tab{Tsummary_ramp} is $2.4\times10^{-7}$ and corresponds
therefore to $h_0^2\OmGW=4\times10^{-12}$.
Likewise, the values of $\hrms^{\rm sat}$ in \Tab{Tsummary_ramp}
have to be multiplied by $a_0^{-1}=8.0\times10^{-16}$ to
obtain the observable $h_c$ at the present time.
Again, the largest value of $\hrms^{\rm sat}=5.1\times10^{-6}$
corresponds therefore to the observable $h_c=4\times10^{-21}$.

\begin{table}[b]\caption{
Characteristic parameters of the runs of Fig.~\ref{pe_comp}.
}\vspace{12pt}\centerline{\begin{tabular}{cccccc}
Run & ${\cal P}$ & $\EEM^{\max}$ & $\EEK^{\max}$ & $\EEGW^{\rm sat}$ & $\hrms^{\rm sat}$\\
\hline
K0  & $0.01$ & $ 0     $ & $ 0.0080$ & $4.5\times10^{-9}$ & $7.0\times10^{-7}$\\
K01 & $0.31$ & $ 0     $ & $ 0.0082$ & $4.4\times10^{-9}$ & $6.9\times10^{-7}$\\
K03 & $0.73$ & $ 0     $ & $ 0.0085$ & $4.8\times10^{-9}$ & $6.4\times10^{-7}$\\
K05 & $0.88$ & $ 0     $ & $ 0.0091$ & $5.3\times10^{-9}$ & $5.7\times10^{-7}$\\
K1  & $0.95$ & $ 0     $ & $ 0.0114$ & $6.4\times10^{-9}$ & $5.5\times10^{-7}$\\
M0  &$-0.01$ & $ 0.0059$ & $ 0.0020$ & $1.6\times10^{-9}$ & $4.1\times10^{-7}$\\
M01 & $0.57$ & $ 0.0078$ & $ 0.0021$ & $2.0\times10^{-9}$ & $4.6\times10^{-7}$\\
M03 & $0.86$ & $ 0.0119$ & $ 0.0024$ & $3.5\times10^{-9}$ & $6.9\times10^{-7}$\\
M05 & $0.94$ & $ 0.0148$ & $ 0.0026$ & $4.5\times10^{-9}$ & $8.6\times10^{-7}$\\
M1  & $0.97$ & $ 0.0168$ & $ 0.0025$ & $5.3\times10^{-9}$ & $9.0\times10^{-7}$\\
\label{Tsummary_pol}\end{tabular}}\end{table}

\subsection{Approach to a Stationary State}

The GW generation can be split into three cases
(see Fig.~\ref{pcomp_M512k2} for kinetically and magnetically driven turbulence):
non-helical, partially helical ($|\sigma|<1$), and fully helical. 
Here, we used $512^3$ meshpoints and arranged the forcing amplitude such
that $\EEK$ and $\EEM$ are around $10^{-2}$.
The viscosity and magnetic diffusivity have equal values and
are chosen such that the dissipative subrange is resolved; see
\Tab{Tsummary_pol} for a summary and Ref.~\cite{DATA} for further details.
Surprisingly, the kinetically driven turbulence is more efficient in producing
GW energy; see Figs.~\ref{pcomp_M512k2}(c) and (f).
However, in this case the presence of kinetic helicity does 
not affect the source amplitude -- contrary to the magnetically 
driven case where the amplitude 
of the source increases substantially with increasing fractional helicity;
see Figs.~\ref{pcomp_M512k2}(a) and (d). 
We also present the kinetic and magnetic energy dissipation rates,
$\epsK$ and $\epsM$, respectively.
The dissipation rates remain almost unchanged during the
stationary stage as we can expect. In addition, 
we see that they are almost unaffected by the presence of helicity.
One may have expected a correlation between $\epsK$ (or $\epsM$)
and $\EEGW$, but in the magnetically driven case, 
larger values of $\sigma$ produce even slightly less dissipation at
early times.
Nevertheless, $\EEGW$ clearly increases with $\sigma$.

\subsection{Polarization Degree}

To estimate the polarization degree, we follow the procedure 
described in Ref.~\cite{Pol:2018pao}. 
We use the usual circular polarization basis tensors
$e^{\pm}_{ij} = -({\bf e}_1 \pm i{\bf e}_2)_i \times ({\bf e}_1 \pm i {\bf e}_2)_j/\sqrt{2}$,
we decompose the Fourier transform of the GW strains 
$h_{ij}({\bf k}) = \int d^3x \, e^{i{\bf k}\cdot {\bf x}} h_{ij}({\bf x})$
into two states -- right- ($h_+$) and left-handed ($h_-$) 
circularly polarized GWs $h_{ij}=h_+e^+_{ij} + h_- e^-_{ij}$.
The GW circular polarization degree is given by
\cite{Kahniashvili:2005qi}\footnote{As an alternative we can use
the decomposition using the {\it linear} polarization basis tensors
$e^+_{ij}({\bf \hat k})=e_i^1 e_j^1 -e_i^2 e_j^2$ and
$e^\times_{ij}({\bf \hat k})=e_i^1 e_j^2 + e_i^2 e_j^1$ as
$h_{ij}({\bf k}) = h_+({\bf k}) e_{ij}^+({\bf \hat k}) + h_\times({\bf k}) e_{ij}^\times({\bf \hat k})$
where $h_+$ and $h_\times$ are gauge independent components corresponding
to two polarization modes.
In this case the polarization degree will be equal to
${\mathcal P}(k) =   {\langle h^{\star}_\times ({\mathbf k})
h_{+}({\mathbf k'}) -
 h^{\star}_+({\mathbf k}) h_\times({\mathbf k'}) \rangle}/
{\langle h^\star_+({\mathbf k}) h_+({\mathbf k'}) +
 h^{\star}_\times({\mathbf k}) h_\times({\mathbf k'}) \rangle}$
} 
\begin{equation}
{\mathcal P}(k) =  \frac {\langle h^{\star}_+({\mathbf k})
h_{+}({\mathbf k'}) -
 h^{\star}_-({\mathbf k}) h_{-}({\mathbf k'}) \rangle}
{\langle h^{\star}_+({\mathbf k}) h_{+}({\mathbf k'}) +
 h^{\star}_-({\mathbf k}) h_{-}({\mathbf k'}) \rangle}
=\frac{{\mathcal H}(k)}{H(k)},~ \label{degree}
\end{equation}
where $H ({k})$ and ${\mathcal H}({k})$ characterize 
the GW amplitude and polarization (chirality) defined through the 
Gaussian-distributed GWs wave number-space two-point 
function (for simplicity of notations we omit the time-dependence): 
$$
\frac{\langle h^{\star}_{ij}({\mathbf k}) h_{lm} ({\mathbf k'})\rangle}
{(2\pi)^3} = \delta^{(3)}({\bf k}-{\bf k'}) \left[ {\mathcal
M}_{ijlm} H(k) + i{\mathcal A}_{ijlm} {\mathcal H} (k) \right], 
$$
where $4 {\mathcal M}_{ijlm} ({\mathbf{\hat k}}) \equiv
P_{il}P_{jm}+P_{im}P_{jl}-P_{ij}P_{lm}$, and
 $8 {\mathcal A}_{ijlm}({\mathbf{\hat k}})
\equiv {\hat {\bf k}}_q (P_{jm} \epsilon_{ilq} + P_{il}
\epsilon_{jmq} + P_{im} \epsilon_{jlq} + P_{jl} \epsilon_{imq})$ are tensors,
$P_{ij}({\bf \hat k}) \equiv \delta_{ij} - {\hat k}_i{\hat k}_j$
is the projection operator, $\delta_{ij}$ and $\epsilon_{ijl}$ are the
Kronecker delta and the fully antisymmetric tensor, respectively.

The GW polarization degree depends, as expected,
on the fractional helicity of the source.
On the other hand, the evolution of helical sources is determined by the
initial fractional helicity: the coupling between the partially helical magnetic
field and the plasma motions leads to a reconfiguration of the magnetic
field at large scales through free decay, resulting in the growth of
the fractional helicity due to the increase of the correlation length
and the corresponding decrease of the magnetic energy until the fully
helical stage is developed and inverse cascading starts
\cite{Tevzadze:2012kk}.
However, for weakly helical sources, a substantial time-period is needed
to reach a fully helical configuration.
On the other hand, our simulations show that the dominant contribution
to the GW signal occurs shortly after the source has reached its maximum:
the subsequent decay of the magnetic field causes a
decline of the turbulent driving of GWs.
This decline is further enhanced by the expansion of the Universe,
although this effect is small if the decay time of the turbulence is 
short compared with the Hubble time; see \Fig{pe_comp}. 
We see that, even with a substantially extended decay phase of the
turbulence of twice the Hubble time, the final GW production is 
enhanced by only a factor of about four.
Thus, the GW polarization will retain information about the
{\it initial} fractional helicity of the source.

\begin{figure}[t!]\begin{center}
\includegraphics[width=\columnwidth]{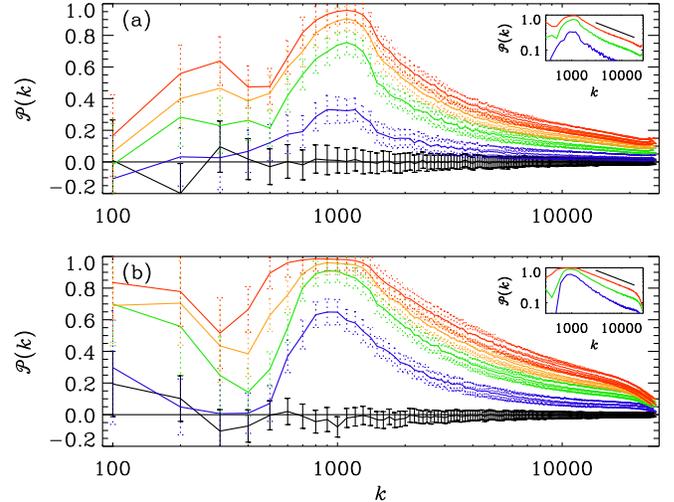}
\end{center}
\caption[]{\small 
Degree of circular polarization for (a) kinetically
and (b) magnetically forced cases with $\sigma=0$ (black)
$0.1$ (blue), $0.3$ (green), $0.5$ (orange), and $1$ (red). 
Approximate error bars based on the temporal fluctuations and
statistical spread for different random seeds of the forcing
are shown as solid black lines for
$\sigma=0$ and as dotted lines otherwise.
The wave number is in units of the comoving Hubble frequency.
}\label{pgrah_comp}\end{figure}

\begin{figure*}[t!]\begin{center}
\includegraphics[width=\textwidth]{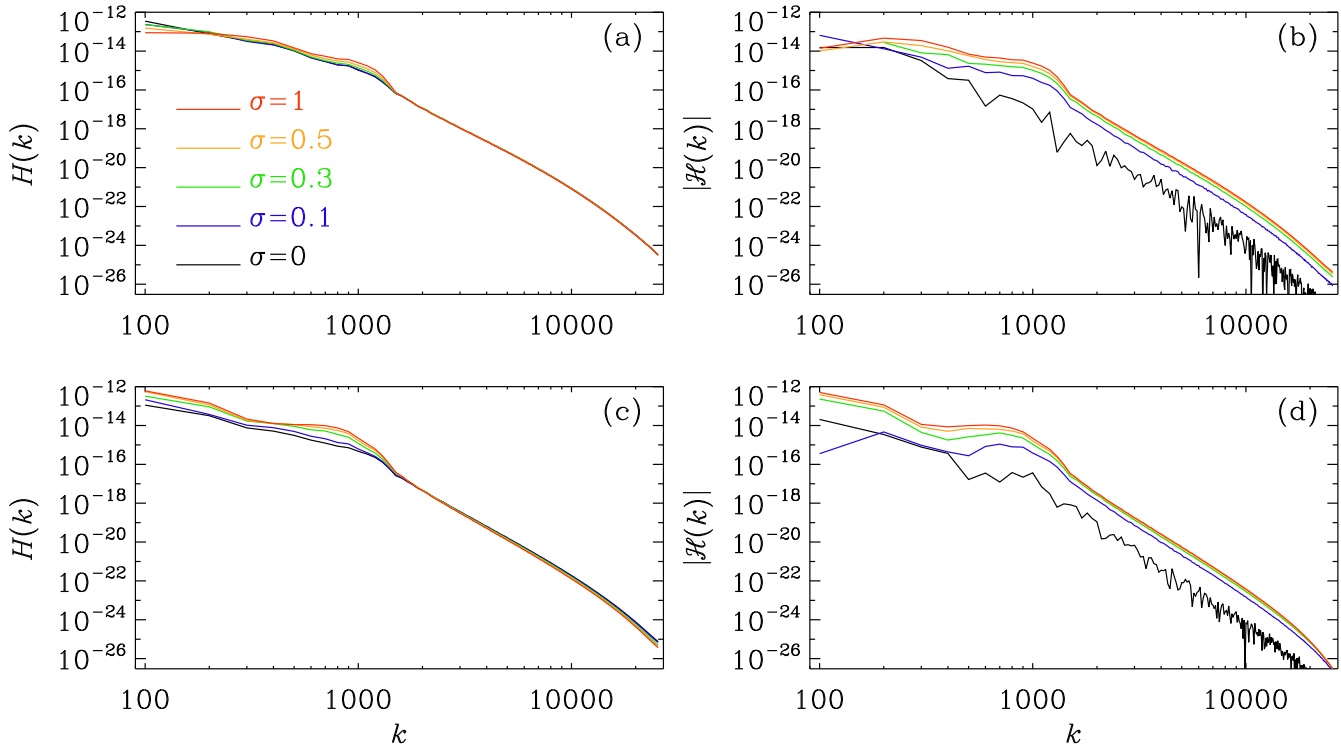}
\end{center}\caption[]{
$H(k)$ (left) and $|{\cal H}(k)|$ (right)
for (a,b) kinetically and (c,d) magnetically forced cases
with $\sigma=0$ (Runs~K0 and M0, black), $0.1$ (Runs~K01 and M01, blue),
$0.3$ (Runs~K03 and M03, green), $0.5$ (Runs~K05 and M05, orange),
and $1$ (Runs~K1 and M1, red).
}\label{pgrah_spec_comp}\end{figure*}

\begin{figure*}[t!]\begin{center}
\includegraphics[width=\textwidth]{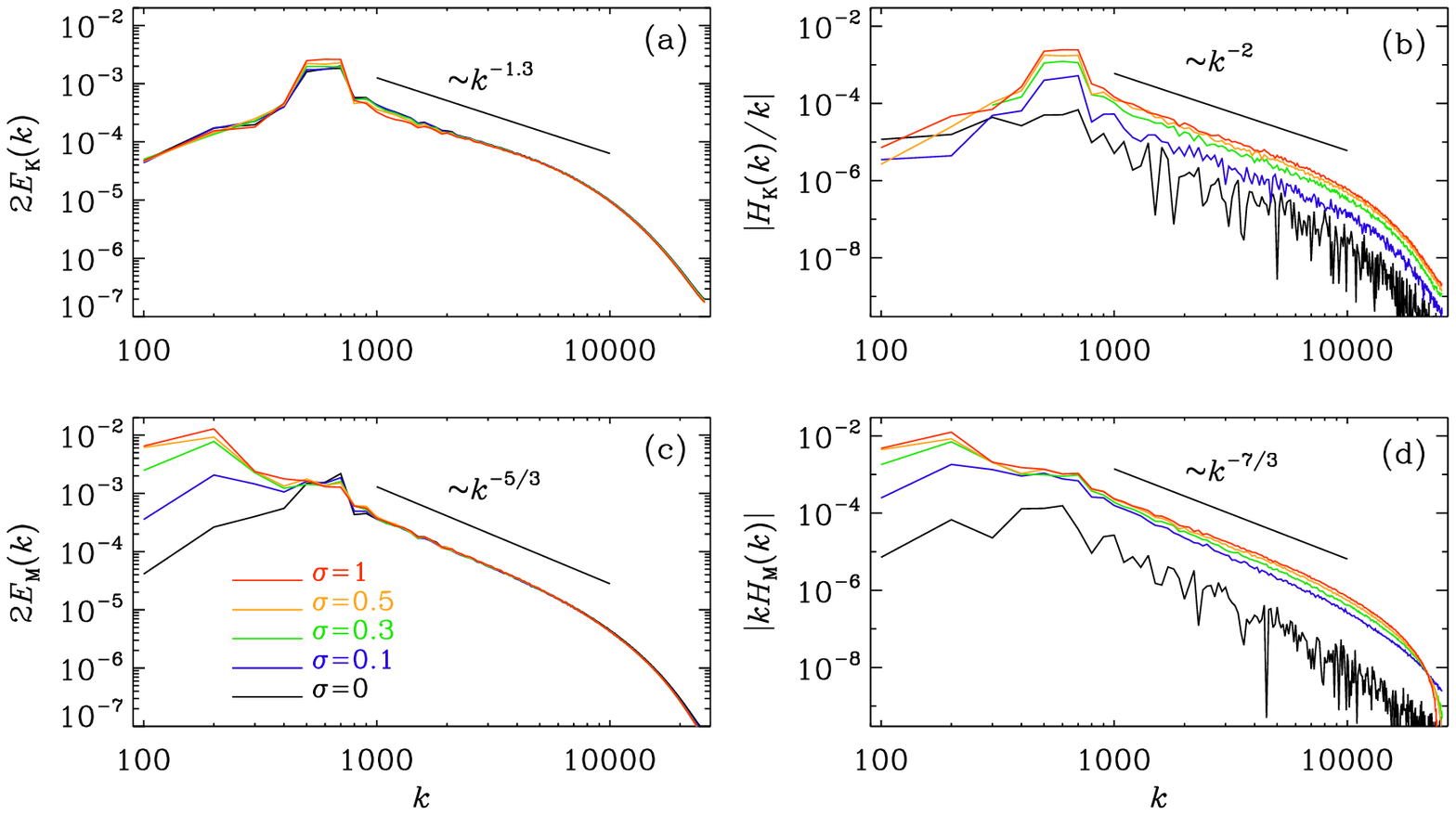}
\end{center}\caption[]{\small 
(a) $2E_K(k)$ and (b) $|H_K(k)/k|$ 
for kinetically forced cases, and
(c) $2E_M(k)$, and (d) $|kH_M(k)|$
for magnetically forced cases
with $\sigma=0$ (Runs~K0 and M0, black), $0.1$ (Runs~K01 and M01, blue),
$0.3$ (Runs~K03 and M03, green), $0.5$ (Runs~K05 and M05, orange),
and $1$ (Runs~K1 and M1, red).
}\label{phel_spec_comp}\end{figure*}

As we have highlighted above, previous works
\cite{Kahniashvili:2005qi,Kisslinger:2015hua} considered stationary
turbulence with two different models for helical turbulence realization:
(i) Kolmogorov-like helical turbulence with two different spectral 
slopes, ${-5/3}$ and $-11/3$ for the spectral energy 
and helicity densities, $E_{\rm M}(k)$ and $H_{\rm M}(k)$, respectively,
and (ii) helicity-transferring
turbulence (if helicity transfer and small-scale helicity dissipation
dominates \cite{Kraichanan:1973}) with the spectral indices 
$-7/3$ and $-10/3$ for $E_{\rm M}(k)$ and $H_{\rm M}(k)$ 
\cite{Moiseev:1996}\footnote{Note that Ref.~\cite{Kahniashvili:2005qi}
used a different convention for the spectral indices referring to the
power spectra of the symmetric $[P_S(k) \propto E_{\rm M}(k)/k^2]$
and the antisymmetric parts $[P_A(k) \propto H_{\rm M}(k)/k]$.}.
The former case seems most suitable for describing the usual nonhelical
turbulence experiencing forward cascading \cite{Borue:1997}.
The difference in using these spectral shapes is determined by the 
effect of helicity on the energy dissipation length.
For highly helical turbulence, the helicity dissipation length 
is larger, so the two region description \cite{Kahniashvili:2005qi}
might be justified using Kolmogorov-like turbulence at large
wavenumbers and approximating the low wave number tail with helicity
transfer turbulence \cite{Ditlevsen:2001}. 
Following this description, Ref.~\cite{Kahniashvili:2008er} and later also 
Ref.~\cite{Ellis:2020uid} discussed two stages --
first the fractionally helical one and later the fully helical one with 
inverse transfer to describe more precisely the GW generation by
helical MHD turbulence.
However, these estimates suffer due to (i) the assumption of stationary
turbulence; (ii) neglecting the decay and correspondingly temporal
dynamical effects (the GW spectrum becomes stationary shortly after
source activation).

\subsection{Dependence on the Driving}

We now investigate the GW polarization spectra dependence on the
nature of driving.
We show the polarization degree spectra in \Fig{pgrah_comp} for 
continuous pumping of kinetic or magnetic energy and 
helicity at intermediate scales.
The parameters of these simulations are the same as
those used for Fig.~\ref{pcomp_M512k2}; see also Ref.~\cite{DATA}.
We see a substantial difference between the kinetic and magnetic 
initial sources at the low frequency tail due to the inverse cascade 
for the magnetic sources that is absent in the kinetically driven case: 
more precisely the energy density spectrum is unchanged after the decay 
stage starts, while the transfer of magnetic helicity to the large scales
results in the increasing of the polarization degree. 
These results confirm that the polarization degree 
is scale dependent: $\propto k^{-0.5}$ 
at large wavenumbers, which is shallower than the $k^{-1}$
expected for Kolmogorov-like helical turbulence with different
spectral indices for the magnetic spectral energy density ($n_S=-5/3$)
and the spectral helicity density ($n_H=-11/3$). 
In our simulations, the actual indices are a bit smaller,
which also explains the shallower slope in the polarization degree.

\subsection{Comparison of the Spectra}

In \Fig{pgrah_spec_comp} we show the numerator and denominator
of the degree of polarization, $H(k)$ and ${\cal H}(k)$, respectively.
The underlying kinetic and magnetic energy and helicity spectra are
shown in \Fig{phel_spec_comp}.
The figure shows that the GW polarization is larger at larger length scales.
To see whether this could be related to inverse cascading in the
magnetic case, we now show in \Fig{phel_spec_comp} the corresponding
energy and helicity spectra.
They show clear inverse cascading of the magnetic
energy and helicity spectra in the magnetically driven case.
Inverse cascading is clearly absent in the kinetic energy and helicity
spectra in the kinetically driven case.

The departure from the theoretical predictions is 
due to the assumption of a scale-independent 
time-decorrelation function 
for magnetic energy and helicity densities. 
Interestingly, the spectral shape of the polarization degree is 
independent of the actual indices for energy and helicity, but depends
on the difference between them 
\cite{Kahniashvili:2005qi}. 
Obviously, in a realistic case, the proper consideration of
time-decorrelation and its dependence on wave numbers is required.
In fact, even in the simplified description, different forms of the
time-decorrelation function for different models of turbulence (including
both compressible and incompressible cases) and its scale-dependence
leads to different scaling of the GW spectrum at high frequencies
\cite{Niksa:2018ofa}.

\section{Conclusions}
In this paper we present numerical simulations of the
circular polarization degree of GWs generated through parity
violating (helical) turbulent sources in the early Universe.
We present our results for GWs generated at electroweak energy scales,
but the formalism is not limited to the specific moment
of GW generation, 
and can be adjusted to primordial turbulence sources 
at any time after inflation and before recombination epochs.  
We have confirmed that the GW signal reaches its maximal strength
faster then the turbulence decays. 
We have also shown that the slope of the 
low frequency tail 
of the GW spectrum
is independent of the nature of the turbulent source (i.e., the 
nature of initial driving, the presence of helicity, etc). 
This restricts the
discrimination between helical and non-helical sources 
as well between kinetic and magnetic drivings
if the signal will be detected.
On the other hand, the polarization spectrum not only 
retains information about 
the initial characteristic frequency (as well as 
the GW signal does) and the strength of parity violation of the source,
but also manifests the dependence in the driving mechanisms.  
We have shown that the previously used assumption of stationary turbulence
does not predict the GW polarization spectrum for realistic turbulence
(the scaling of the spectrum at low and high frequencies, the peak
position, etc), and thus might result in inadequate estimates of the
detection prospects. 
In particular, we have shown that the polarization spectra increase at 
low frequencies due to the inverse cascade,
reflecting redistribution of the helical structures, while the energy
density of the GWs sustains the scaling that was established soon after
the turbulent source activation. 
In fact, previous works \cite{Kahniashvili:2005qi,Ellis:2020uid} 
predicted a completely different picture for both helical Kolmogorov 
turbulence and helicity-transferring turbulence. 
More precisely, no previous work addressed the dependence of the
polarization spectra on the temporal characteristics of the source, 
nor was the helicity transfer-induced increase of polarization 
spectra at low frequencies detected. There was never a 
self-consistent description of GW polarization nor an attempt to disentangle 
the nature of the underlying source (kinetic vs magnetic driving).
Based on our results, we argue that the inverse cascade time scale determines 
the height of the polarization spectra at low frequencies (a second peak) 
in the MHD case -- even for low fractional helicity.
The second (smaller) peak is located at lower frequencies.
Fortunately, numerical simulations have now become an affordable tool
to address these and other questions of relic GW 
generation 
and give a more complete picture for the detection prospects by 
LISA (as for electroweak phase transitions) and by PTAs (as for 
QCD phase transitions 
\cite{Caprini:2010xv,Neronov:2020qrl}), and/or any future planned
missions, including atomic interferometry \cite{Dimopoulos:2007cj}.

\vspace{2mm}
Data availability---The source code used for the
simulations of this study, the {\sc Pencil Code},
is freely available from Ref.~\cite{PC}.
The simulation setups and the corresponding data
are freely available from Ref.~\cite{DATA}.

\acknowledgements 
We thank Arthur Kosowsky and Andrii Neronov 
for useful discussions. 
Support through the Swedish Research Council, grant 2019-04234,
and Shota Rustaveli GNSF  (grants FR/18-1462 and FR/19-8306)
are gratefully acknowledged.
We acknowledge the allocation of computing resources provided by the
Swedish National Allocations Committee at the Center for Parallel
Computers at the Royal Institute of Technology in Stockholm.

\end{document}